\documentclass[aps,twocolumn,prd,preprint,tightenlines,superscriptaddress,showpacs,byrevtex]{revtex4}
\usepackage{easy-todo}
\usepackage{overpic}
\usepackage{graphicx}
\usepackage{dcolumn}
\usepackage{bm}
\usepackage{xcolor}
\usepackage{amsmath}

\begin{document}
\preprint{\vbox{ \hbox{   }
    \hbox{BELLE Preprint  {\it 2017-26}}
    \hbox{KEK Preprint  {\it 2017-39}}
}}
\noaffiliation
\affiliation{University of the Basque Country UPV/EHU, 48080 Bilbao}
\affiliation{Beihang University, Beijing 100191}
\affiliation{Budker Institute of Nuclear Physics SB RAS, Novosibirsk 630090}
\affiliation{Faculty of Mathematics and Physics, Charles University, 121 16 Prague}
\affiliation{Chonnam National University, Kwangju 660-701}
\affiliation{University of Cincinnati, Cincinnati, Ohio 45221}
\affiliation{Deutsches Elektronen--Synchrotron, 22607 Hamburg}
\affiliation{University of Florida, Gainesville, Florida 32611}
\affiliation{Gifu University, Gifu 501-1193}
\affiliation{SOKENDAI (The Graduate University for Advanced Studies), Hayama 240-0193}
\affiliation{Hanyang University, Seoul 133-791}
\affiliation{University of Hawaii, Honolulu, Hawaii 96822}
\affiliation{High Energy Accelerator Research Organization (KEK), Tsukuba 305-0801}
\affiliation{J-PARC Branch, KEK Theory Center, High Energy Accelerator Research Organization (KEK), Tsukuba 305-0801}
\affiliation{IKERBASQUE, Basque Foundation for Science, 48013 Bilbao}
\affiliation{Indian Institute of Science Education and Research Mohali, SAS Nagar, 140306}
\affiliation{Indian Institute of Technology Bhubaneswar, Satya Nagar 751007}
\affiliation{Indian Institute of Technology Guwahati, Assam 781039}
\affiliation{Indian Institute of Technology Hyderabad, Telangana 502285}
\affiliation{Indian Institute of Technology Madras, Chennai 600036}
\affiliation{Indiana University, Bloomington, Indiana 47408}
\affiliation{Institute of High Energy Physics, Chinese Academy of Sciences, Beijing 100049}
\affiliation{Institute of High Energy Physics, Vienna 1050}
\affiliation{Institute for High Energy Physics, Protvino 142281}
\affiliation{INFN - Sezione di Napoli, 80126 Napoli}
\affiliation{INFN - Sezione di Torino, 10125 Torino}
\affiliation{Advanced Science Research Center, Japan Atomic Energy Agency, Naka 319-1195}
\affiliation{J. Stefan Institute, 1000 Ljubljana}
\affiliation{Kanagawa University, Yokohama 221-8686}
\affiliation{Institut f\"ur Experimentelle Kernphysik, Karlsruher Institut f\"ur Technologie, 76131 Karlsruhe}
\affiliation{Kennesaw State University, Kennesaw, Georgia 30144}
\affiliation{King Abdulaziz City for Science and Technology, Riyadh 11442}
\affiliation{Department of Physics, Faculty of Science, King Abdulaziz University, Jeddah 21589}
\affiliation{Korea Institute of Science and Technology Information, Daejeon 305-806}
\affiliation{Korea University, Seoul 136-713}
\affiliation{Kyoto University, Kyoto 606-8502}
\affiliation{Kyungpook National University, Daegu 702-701}
\affiliation{\'Ecole Polytechnique F\'ed\'erale de Lausanne (EPFL), Lausanne 1015}
\affiliation{P.N. Lebedev Physical Institute of the Russian Academy of Sciences, Moscow 119991}
\affiliation{Faculty of Mathematics and Physics, University of Ljubljana, 1000 Ljubljana}
\affiliation{Ludwig Maximilians University, 80539 Munich}
\affiliation{Luther College, Decorah, Iowa 52101}
\affiliation{University of Malaya, 50603 Kuala Lumpur}
\affiliation{University of Maribor, 2000 Maribor}
\affiliation{Max-Planck-Institut f\"ur Physik, 80805 M\"unchen}
\affiliation{School of Physics, University of Melbourne, Victoria 3010}
\affiliation{University of Mississippi, University, Mississippi 38677}
\affiliation{University of Miyazaki, Miyazaki 889-2192}
\affiliation{Moscow Physical Engineering Institute, Moscow 115409}
\affiliation{Moscow Institute of Physics and Technology, Moscow Region 141700}
\affiliation{Graduate School of Science, Nagoya University, Nagoya 464-8602}
\affiliation{Kobayashi-Maskawa Institute, Nagoya University, Nagoya 464-8602}
\affiliation{Nara Women's University, Nara 630-8506}
\affiliation{National Central University, Chung-li 32054}
\affiliation{National United University, Miao Li 36003}
\affiliation{Department of Physics, National Taiwan University, Taipei 10617}
\affiliation{H. Niewodniczanski Institute of Nuclear Physics, Krakow 31-342}
\affiliation{Nippon Dental University, Niigata 951-8580}
\affiliation{Niigata University, Niigata 950-2181}
\affiliation{Novosibirsk State University, Novosibirsk 630090}
\affiliation{Osaka City University, Osaka 558-8585}
\affiliation{Pacific Northwest National Laboratory, Richland, Washington 99352}
\affiliation{Panjab University, Chandigarh 160014}
\affiliation{University of Pittsburgh, Pittsburgh, Pennsylvania 15260}
\affiliation{Research Center for Nuclear Physics, Osaka University, Osaka 567-0047}
\affiliation{Theoretical Research Division, Nishina Center, RIKEN, Saitama 351-0198}
\affiliation{University of Science and Technology of China, Hefei 230026}
\affiliation{Showa Pharmaceutical University, Tokyo 194-8543}
\affiliation{Soongsil University, Seoul 156-743}
\affiliation{University of South Carolina, Columbia, South Carolina 29208}
\affiliation{Stefan Meyer Institute for Subatomic Physics, Vienna 1090}
\affiliation{Sungkyunkwan University, Suwon 440-746}
\affiliation{School of Physics, University of Sydney, New South Wales 2006}
\affiliation{Department of Physics, Faculty of Science, University of Tabuk, Tabuk 71451}
\affiliation{Tata Institute of Fundamental Research, Mumbai 400005}
\affiliation{Excellence Cluster Universe, Technische Universit\"at M\"unchen, 85748 Garching}
\affiliation{Department of Physics, Technische Universit\"at M\"unchen, 85748 Garching}
\affiliation{Department of Physics, Tohoku University, Sendai 980-8578}
\affiliation{Earthquake Research Institute, University of Tokyo, Tokyo 113-0032}
\affiliation{Department of Physics, University of Tokyo, Tokyo 113-0033}
\affiliation{Tokyo Institute of Technology, Tokyo 152-8550}
\affiliation{Tokyo Metropolitan University, Tokyo 192-0397}
\affiliation{University of Torino, 10124 Torino}
\affiliation{Virginia Polytechnic Institute and State University, Blacksburg, Virginia 24061}
\affiliation{Wayne State University, Detroit, Michigan 48202}
\affiliation{Yamagata University, Yamagata 990-8560}
\affiliation{Yonsei University, Seoul 120-749}
  \author{M.~Berger}\affiliation{Stefan Meyer Institute for Subatomic Physics, Vienna 1090} 
  \author{C.~Schwanda}\affiliation{Institute of High Energy Physics, Vienna 1050} 
  \author{K.~Suzuki}\affiliation{Stefan Meyer Institute for Subatomic Physics, Vienna 1090}
  \author{I.~Adachi}\affiliation{High Energy Accelerator Research Organization (KEK), Tsukuba 305-0801}\affiliation{SOKENDAI (The Graduate University for Advanced Studies), Hayama 240-0193} 
  \author{J.~K.~Ahn}\affiliation{Korea University, Seoul 136-713} 
  \author{H.~Aihara}\affiliation{Department of Physics, University of Tokyo, Tokyo 113-0033} 
  \author{S.~Al~Said}\affiliation{Department of Physics, Faculty of Science, University of Tabuk, Tabuk 71451}\affiliation{Department of Physics, Faculty of Science, King Abdulaziz University, Jeddah 21589} 
  \author{D.~M.~Asner}\affiliation{Pacific Northwest National Laboratory, Richland, Washington 99352} 
  \author{H.~Atmacan}\affiliation{University of South Carolina, Columbia, South Carolina 29208} 
  \author{V.~Aulchenko}\affiliation{Budker Institute of Nuclear Physics SB RAS, Novosibirsk 630090}\affiliation{Novosibirsk State University, Novosibirsk 630090} 
% ...
\collaboration{The Belle Collaboration}

\title{\quad\\[1.0cm]Measurement of the Decays $\Lambda_c\to \Sigma\pi\pi$ at Belle}


\date{\today}

\begin{abstract}
We report measurements of the branching fractions of the decays
$\Lambda^+_c\to\Sigma^+\pi^-\pi^+$, $\Lambda^+_c\to\Sigma^0\pi^+\pi^0$
and $\Lambda^+_c\to\Sigma^+\pi^0\pi^0$ relative to the reference channel $\Lambda^+_c\to pK^-\pi^+$. The analysis is based on the full data sample collected at and near the $\Upsilon(4S)$ resonance by the Belle detector at the KEKB asymmetric-energy $e^+e^-$ collider, corresponding to an integrated luminosity of 711 fb$^{-1}$. We measure ${\cal B}(\Lambda^+_c\rightarrow\Sigma^+\pi^-\pi^+)/{\cal B}(\Lambda^+_c\to pK^-\pi^+) = 0.7{19}~\pm 0.003~\pm 0.02{4}$,
${\cal B}(\Lambda^+_c\rightarrow\Sigma^0\pi^+\pi^0)/{\cal B}(\Lambda^+_c\to pK^-\pi^+) =0.{575}~\pm 0.005~\pm 0.0{36}$ and
${\cal B}(\Lambda^+_c\rightarrow\Sigma^+\pi^0\pi^0)/{\cal B}(\Lambda^+_c\to pK^-\pi^+) = 0.{247}~\pm 0.006~\pm 0.01{9}$. The listed uncertainties are statistical and systematic, respectively.
\end{abstract}

\maketitle


\section{Introduction} \label{sec:mot}

Charmed baryon decays provide crucial information for the study of both strong and weak interactions. The $\Lambda_c$, which is the lightest charmed baryon and has a $udc$ quark configuration, plays a key role. 
As most $\Lambda_b^0$ decays include a $\Lambda_c^+$~\cite{Lambda_b1,Lambda_b2} in their decay products, improved measurements of $\Lambda_c^+$ hadronic branching fractions help constrain fragmentation functions of bottom, as well as charm, quarks through the measurement of inclusive heavy-flavor baryon production~\cite{fracfunc1}~\cite{fracfunc2}.
The recent model-independent measurements of the normalization mode $\Lambda_c\to pK\pi$ by Belle~\cite{pkpi_Belle} and BESIII~\cite{pkpi_BESIII} improve the accuracy of $\Lambda_c^+$ branching fractions measured relative to this mode and similarly advance other related measurements~\cite{wrongpkpi}. 
The decay $\Lambda_c^+\to\Sigma\pi\pi$ is particularly interesting as it 
has been proposed as a possible avenue to extract the $\Sigma $-$\pi$ scattering length~\cite{HO11}, and this measurement would provide crucial information in the study of the $\Lambda(1405)$ resonance~\cite{lambda1405}.

In this paper, we report measurements of the branching fractions of the decays $\Lambda^+_c\to\Sigma^+\pi^-\pi^+$, $\Lambda^+_c\to\Sigma^0\pi^+\pi^0$ and $\Lambda^+_c\to\Sigma^+\pi^0\pi^0$ relative to the reference channel $\Lambda^+_c\to pK^-\pi^+$~\cite{cc}.

This analysis is based on the full Belle data sample taken at the $\Upsilon(4S)$ resonance. In principle, it would be desirable to also measure $\Lambda^+_c\to\Sigma^-\pi^+\pi^+$. However $\Sigma^-$ decays almost completely into $n \pi^-$, a mode that cannot be reconstructed at Belle. Belle's inability to measure neutrons also limits us to the decay modes $\Sigma^+\to p \pi^0$ and $\Lambda \to p \pi^-$ when reconstructing hyperons. While the $\Lambda^+_c\to\Sigma^+\pi^-\pi^+$ and $\Lambda^+_c\to\Sigma^0\pi^+\pi^0$ modes have been studied previously by BESIII~\cite{pkpi_BESIII} and by CLEO~\cite{CLEO_Sig0}, respectively, we present here the first measurement of the $\Lambda^+_c\to\Sigma^+\pi^0\pi^0$ channel.

\section{Experimental procedure}

\subsection{Data sample} \label{sec:sample}

This analysis is based on the 711~fb$^{-1}$ data sample collected with the Belle detector at the KEKB asymmetric-energy $e^+e^-$ collider~\cite{KEKB} operating at an energy at or near the $\Upsilon(4S)$ resonance. Belle is a large-solid-angle magnetic spectrometer that consists of a silicon vertex detector (SVD), a 50-layer central drift chamber (CDC), an array of
aerogel threshold Cherenkov counters (ACC),  
a barrel-like arrangement of time-of-flight scintillation counters (TOF), and an electromagnetic calorimeter comprised of CsI(Tl) crystals (ECL) located inside a superconducting solenoid coil that provides a 1.5~T magnetic field.  An iron flux return located outside of the coil is instrumented to detect $K_L^0$ mesons and to identify muons (KLM).
Two inner detector configurations were used. A 2.0~cm radius beampipe
and a 3-layer silicon vertex detector were used for the first sample
of 140~fb$^{-1}$, while a 1.5~cm radius beampipe, a 4-layer
silicon detector and a small-cell inner drift chamber were used to record  
the remaining 571~fb$^{-1}$~\cite{svd2}. The detector is described in detail elsewhere~\cite{Belle}.

In addition, we use Monte Carlo (MC) simulated events, which are created with the JETSET~\cite{jetset} and EVTGEN~\cite{evtgen} generators. A full detector simulation based on GEANT3~\cite{geant3} is applied to MC events to model the response of the detector and its acceptance. Final-state radiation is taken into account using the PHOTOS~\cite{Photos} package. MC-simulated data samples are equivalent to at least six times the data luminosity.

\subsection{Event selection} \label{sec:cuts} \label{sec:presel}

Charged particles are reconstructed in the tracking system consisting of the SVD and CDC detectors. Particle identification is based on the specific ionization in the CDC, the Cherenkov light yield in the ACC, and the time-of-flight information in the TOF. For each track, the normalized likelihood ratio for distinct hypotheses $i \in \{p,\ \pi,\ K\}$ and $j \ne i$ is defined as ${\cal L}(i:j)={\cal L}(i)/({\cal L}(i)+{\cal L}(j))$. For a track to be identified as a proton (pion), the corresponding likelihood ratios must exceed 0.6. For $pK^-\pi^+$ alone, the more stringent requirement of ${\cal L}(p:K)>0.9$ and ${\cal L}(p:\pi)>0.9$ for proton candidates is adopted. These selection criteria are about 90\,\% efficient for detected kaons, 98\,\% for pions and 80\,\% (90\,\%) for protons coming directly from $\Lambda_c$ (from hyperons). For all charged particles except the protons and pions from the $\Sigma^+$ and $\Lambda$ decays, we require the distance of closest approach $|dz|$ ($dr$) to the interaction point (IP) along the beam axis (in the transverse plane) to be smaller than 4~cm (2~cm).

Photons are reconstructed from clusters in the ECL are not matched to a CDC track trajectory. We require a minimum cluster energy of 40 MeV. A neutral $\pi^0$ candidate is formed by combining two photons selected in a $M(\gamma\gamma)$ window of [120, 150]~MeV/$c^2$ (about $\pm 3\sigma$  around the nominal $\pi^0$ mass). The reconstructed $\pi^0$~momentum must exceed 100~MeV/$c$ in the laboratory frame.

A $\Lambda$ candidate is reconstructed by combining a proton and a pion with an invariant-mass $M(p\pi)$ between 1.1130 and 1.1180~GeV/$c^2$ (about $\pm 3\sigma$ around the nominal $\Lambda$ mass). In Belle analyses, additional criteria may be applied, based on the distance along the beam axis of the two daughter tracks at their closest approach ($z_{\rm dist}$), the minimum $dr$ of each track, the angular difference in the transverse plane between the $\Lambda$ flight direction and the vector between the IP and the decay vertex ($d\phi$), and the flight length in the transverse plane of the $\Lambda$ candidate ($|\ell_f|$). Two levels of $\Lambda$~candidate purity are commonly used in Belle, based on the selection criteria for these four parameters~\cite{lambda1}\cite{lambda2}\cite{lambda3}. Level 1 (2) is determined by optimizing these $\Lambda$-selection criteria on MC samples after (without) selections on the charged particle likelihood ratios. The threshold values for each parameter are given in Table~\ref{tab:purity} for the two levels. However, at this point we make no selection based on the purity level.
\begin{table}
  \caption{Selection criteria for $\Lambda$ at purity level 1 (level 2) as commonly used in the Belle collaboration.}  \label{tab:purity}
  \centering
  \begin{tabular}{cccc}
    \hline \hline
 $p$~[GeV/$c$] & $<$ 0.5 & 0.5  $-$ 1.5  & $>$ 1.5 \\
    \hline
   Max $z_{\rm dist}$~[cm]&  12.9(7.7) &  9.8(2.1) &  2.4(1.9) \\ 
   Min $dr$~[mm]&  0.08(0.18) &  0.10(0.33) &  0.27(0.59) \\ 
   Max $d\phi$~[${}^{\circ}$]&  0.09(0.07) &  0.18(0.10) &  1.20(0.60) \\ 
   Min $|\ell_f|$~[mm] & 2.2(3.5) & 1.6(2.4) & 1.1(1.7) \\ 
    
    \hline\hline
  \end{tabular}
\end{table}

A $\Sigma^0$ candidate is formed by combining a $\Lambda$ candidate with a photon, with $M(\Lambda\gamma)$ required to lie between {1.18 and 1.206}~GeV/$c^2$ (about $\pm 3\sigma$). Similarly, a $\Sigma^+$ candidate is formed from the combination of a proton with a $\pi^0$, with $M(p\pi^0)$ lying between 1.159 and 1.219 GeV/$c^2$ (about $\pm 2.5\sigma$). The $\Sigma^+\to p\pi^0$ reconstruction relies on the long hyperon lifetime: we require the proton's $dr$ to exceed 0.3~mm. Then, the $\Sigma^+$ trajectory is approximated by a straight line from the IP in the direction of the reconstructed $\Sigma^+$ three-momentum and intersected with the proton path. This point is taken as an estimate of the $\Sigma^+$ decay vertex and used to refit the $\pi^0$~candidate, assuming that the $\gamma\gamma$~pair originates from this vertex rather than from the IP. Only $\Sigma^+$~candidates with a positive flight length from the IP to the decay vertex are retained.

Finally, the $\Sigma$ baryon candidate is combined with two pions. To reduce combinatorial background, the scaled momentum $x = p/p_{\text{max}}$ is required to be larger than 0.5. Here, $p$ is the magnitude of the $\Lambda_c^+$ three-momentum and $p_{\text{max}}$ is its maximum value assuming only a pair of $\Lambda_c^+$ baryons is produced in the event. As a consequence of this requirement, all $\Lambda_c^+$ candidates from $B$~decays are completely eliminated and only candidates originating directly from the $e^+ e^- \to c\bar{c}$ continuum are retained. Charged daughter particles are fitted to a common decay vertex; the $\chi^2$ of this fit is required to be compatible with the daughters being produced by a common parent.

\subsection{Boosted decision tree selector} \label{sec-BDT-input}

To further increase the purity of the reconstructed signal, we combine several discriminant variables into a single Boosted Decision Tree (BDT) output, based on the AdaBoost~\cite{adaboost} algorithm.

The input variables to the BDT are: the scaled momenta of the $\Lambda_c^+$~candidate and the hyperon, all final-state charged-particle and $\pi^0$ candidate momenta in the center-of-mass (c.m.) frame, the cluster energy and direction of detected photons in the ECL, the cosine of the angle between the two photons from all $\pi^0$ particles in the laboratory frame, the $\chi^2$ of the vertex fit (described above) in modes with several charged daughters, the distances of closest approach to the interaction point ($dr$, $|dz|$) of all charged trajectories, the $\Lambda$-candidate purity level (described earlier), and a purity flag for each $\pi^0$ candidate. This binary flag is assigned by (1) forming $\pi^0$ candidates from all possible two-photon combinations, starting from the most energetic photons, then (2) processing this ordered list to assign a value of one for the first combination with an invariant-mass in the range of $\pm 15$~MeV/$c^2$ of the nominal $\pi^0$~mass and zero for all other combinations using the same photons. This requirement ensures that only the most likely $\gamma\gamma$ combinations are used and avoids double counting.

The classifier is trained on MC event samples corresponding to the same integrated luminosity as the real data sample except in the case of the $\Sigma^+\pi^0\pi^0$ decay mode, where six times the real data luminosity is used. If there are multiple candidates in one event, the one with the highest-ranking BDT classifier is selected. The selection threshold applied to the BDT output is optimized by maximizing a figure of merit defined as $S/\sqrt{S+B}$, where $S$ represents the number of signal events and $B$ the number of background events that pass the selection criteria, as estimated from MC samples introduced earlier. For the $\Sigma^+\pi^0\pi^0$ channel, where no previous measurement is available, a branching fraction of 1.8\,\% is assumed from isospin considerations.

\subsection{Signal yield extraction}

\begin{figure}
  \raggedright
  \begin{overpic}[width=0.48\textwidth]{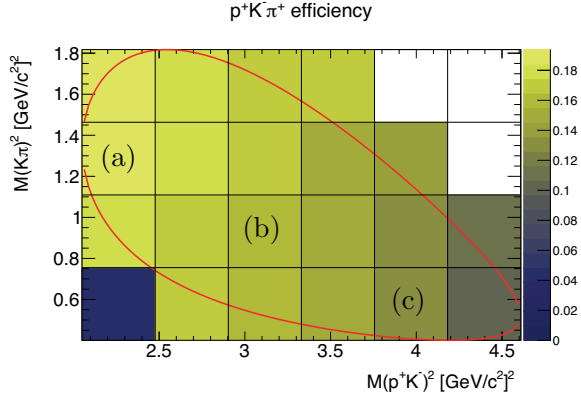}
   \put(16,38){(a)}	
  \put(40,26){(b)}
  \put(65,14){(c)}
  \end{overpic}

  \caption{Dalitz distribution binning and reconstruction efficiency in bins of $M(pK^-)^2$ vs.\ $M(K^-\pi^+)^2$ for the $\Lambda^+_c\to pK^-\pi^+$~channel. The curved line is the kinematic boundary of the Dalitz plot. The fits for yield extraction in bins (a), (b) and (c) are shown in Fig.~\ref{fig:pkpi_fits}.\label{fig:pkpi}}
\end{figure}

\begin{figure}
\includegraphics[width=0.48\textwidth]{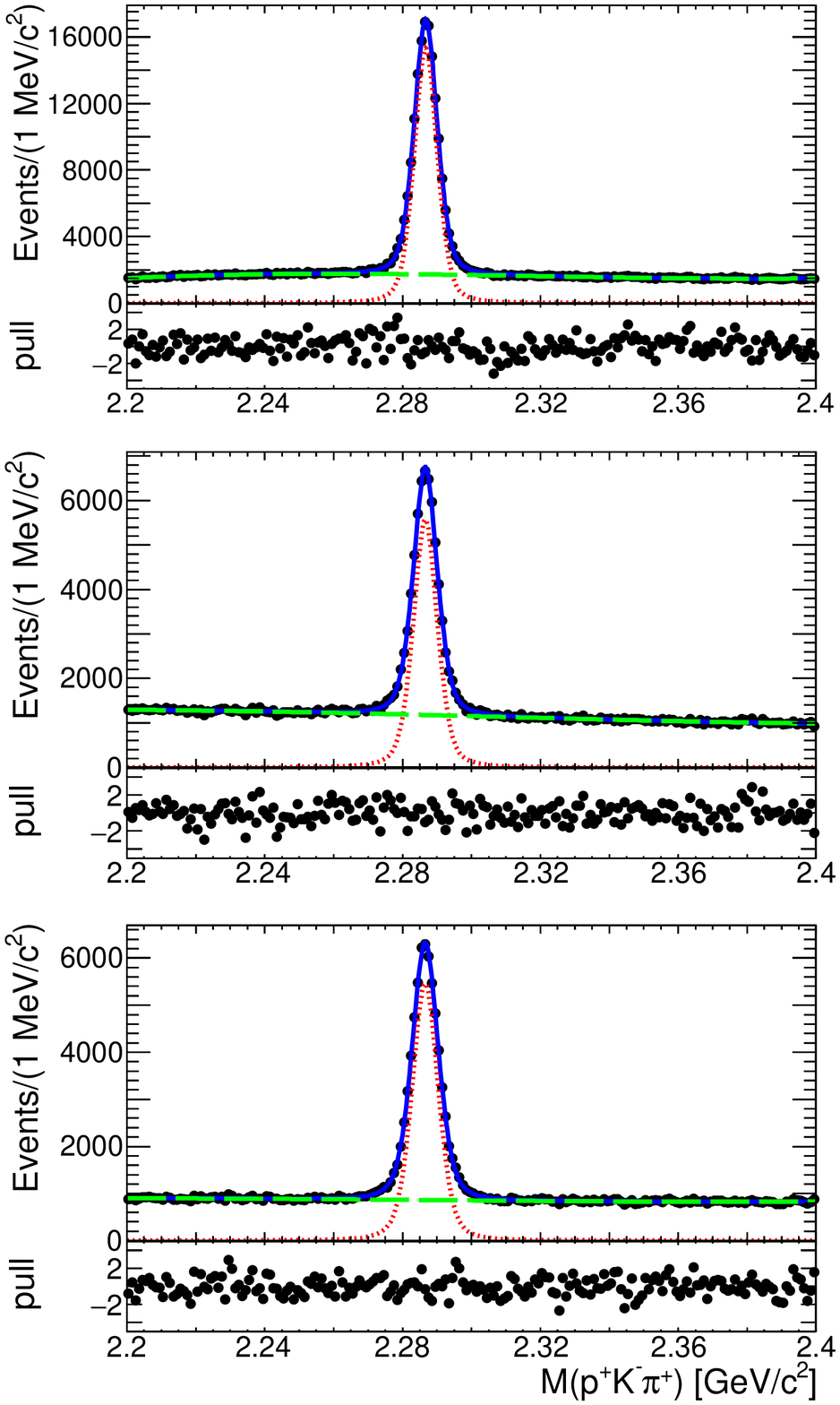}
  \caption{Fits (solid curves) in three representative Dalitz bins of the $\Lambda^+_c\to pK^-\pi^+$~channel. From top to bottom, the panels correspond to bins (a), (b) and (c) in Fig.~\ref{fig:pkpi}. The signal is shown as the dotted curve and the combinatorial background as the dashed curve. The pull distribution of the fit is shown at the bottom of each panel.} \label{fig:pkpi_fits}
\end{figure}

\begin{figure}
  \raggedright
  \begin{overpic}[width=0.48\textwidth]{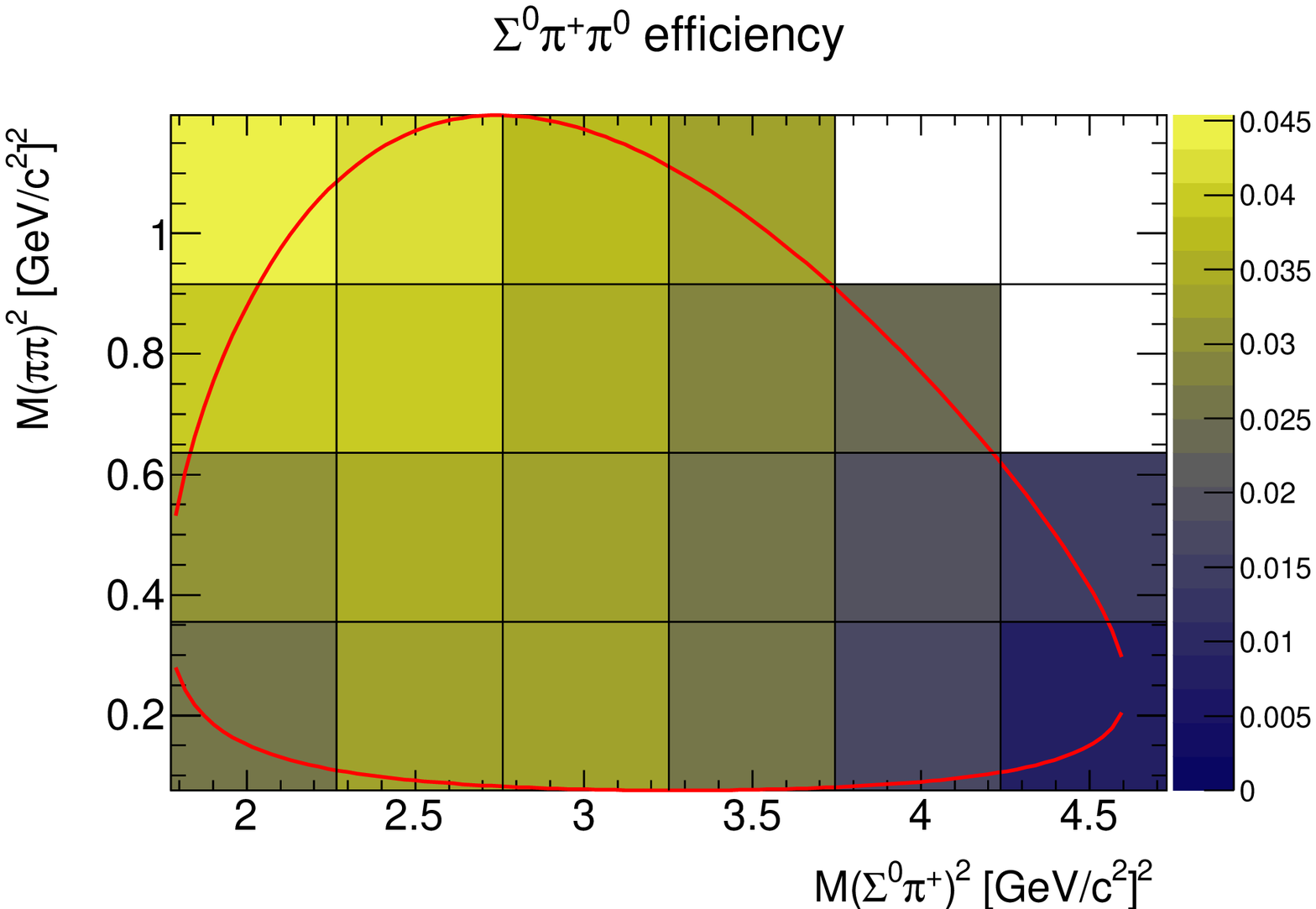}
\put(16,38){(a)}
\put(40,26){(b)}
\put(65,14){\color{white}(c)}
  \end{overpic}

  \caption{Dalitz distribution binning and reconstruction efficiency in bins of $M(\Sigma^0\pi^+)^2$ vs.\ $M(\pi^0\pi^+)^2$ for the $\Lambda^+_c\to\Sigma^0\pi^+\pi^0$~channel. The curved line is the kinematic boundary of the Dalitz plot. The fits in representative bins (a), (b) and (c) are shown in Fig.~\ref{fig:Sig0_fits}.} \label{fig:Sig0}
\end{figure}

\begin{figure}
\includegraphics[width=0.48\textwidth]{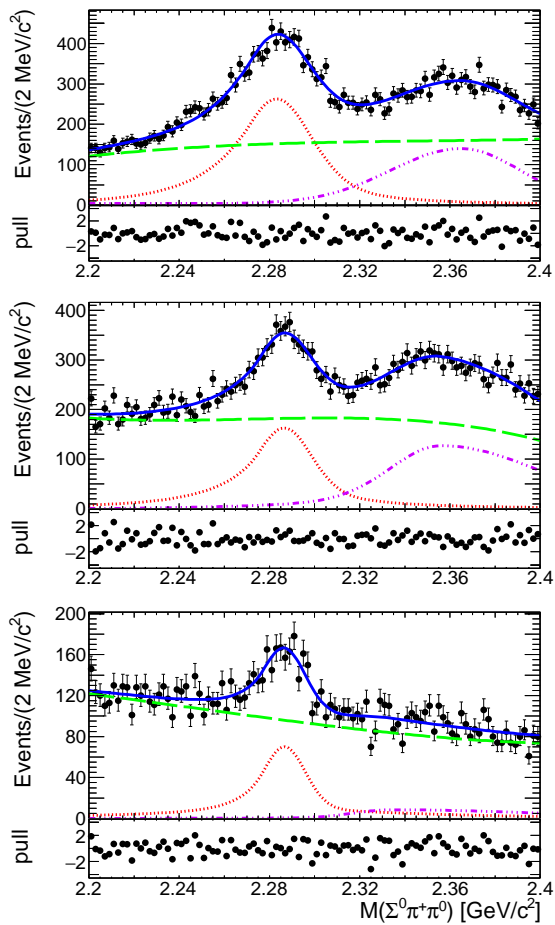}
  \caption{Fits (solid curves) in three representative Dalitz bins of the $\Lambda^+_c\to\Sigma^0\pi^+\pi^0$~channel. From top to bottom, the panels correspond to bins (a), (b) and (c) in Fig.~\ref{fig:Sig0}. The dotted curve is the signal component, the dashed curve the combinatorial background, and the  dash--dotted curve the $\Lambda\pi^+\pi^0+\gamma$ background. The pull distribution of the fit is shown at the bottom of each panel.} \label{fig:Sig0_fits}
\end{figure}

\begin{figure}
  \raggedright
  \begin{overpic}[width=0.48\textwidth]{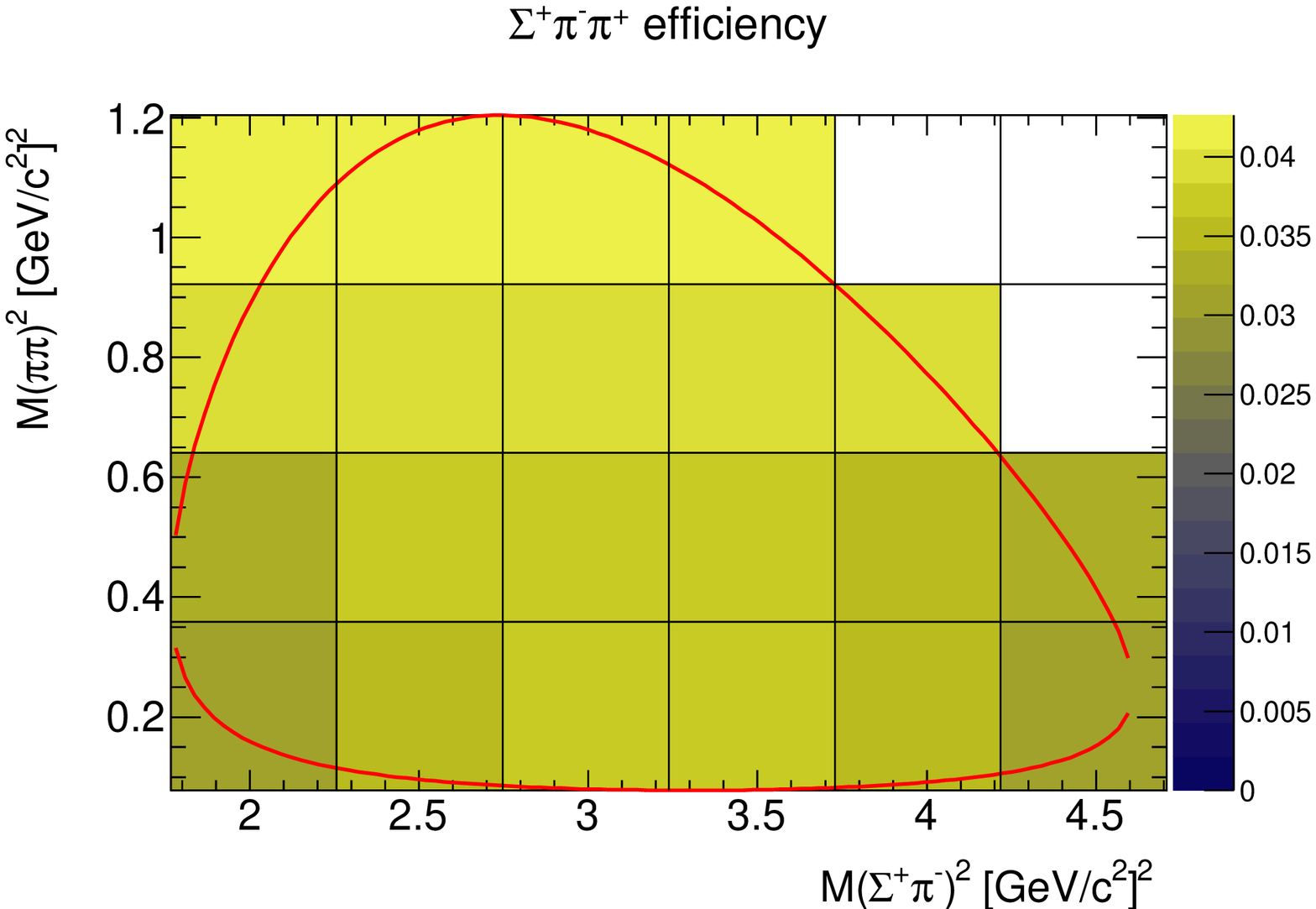}
   \put(15,38){(a)}
   \put(52,38){(b)}
   \put(65,14){(c)}
  \end{overpic}

 \caption{Dalitz distribution binning and reconstruction efficiency in bins of $M(\Sigma^+\pi^-)^2$ vs.\ $M(\pi^-\pi^+)^2$ for the $\Lambda^+_c\to\Sigma^+\pi^-\pi^+$~channel. The curved line is the kinematic boundary of the Dalitz plot. The fit results in representative bins (a), (b) and (c) are shown in Fig.~\ref{fig:Sigp_fits}.} \label{fig:Sigp}
\end{figure}

\begin{figure}
\includegraphics[width=0.48\textwidth]{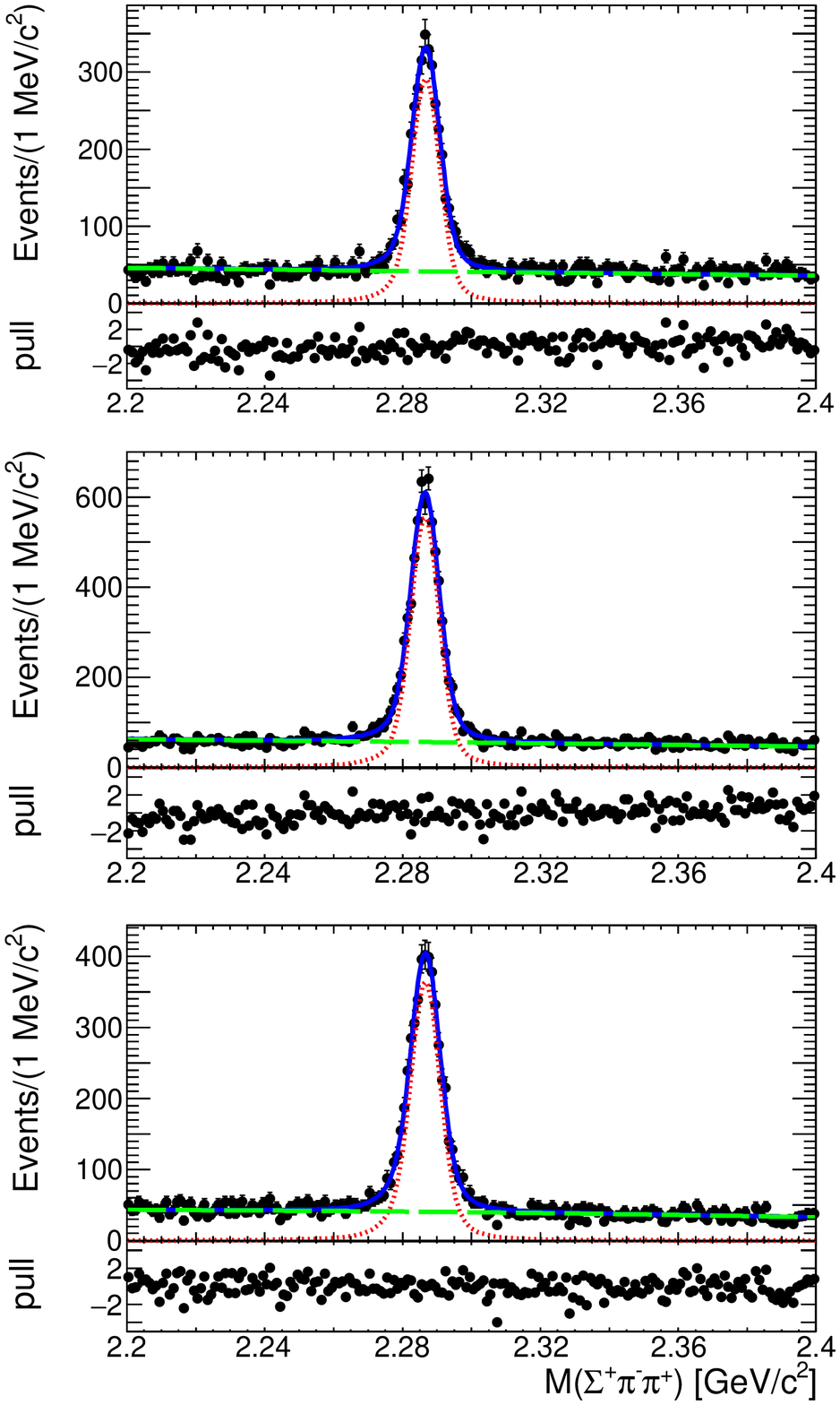}
  \caption{Fits (solid curves) in three representative Dalitz bins of the $\Lambda^+_c\to\Sigma^+\pi^-\pi^+$~channel. From top to bottom, the panels correspond to bins (a), (b) and (c) in Fig.~\ref{fig:Sigp}. The signal component is shown as the dotted curve, the combinatorial background as the dashed curve. The pull distribution of the fit is shown at the bottom of each panel.} \label{fig:Sigp_fits} 
\end{figure}

\begin{figure}
  \raggedright
  \begin{overpic}[width=0.48\textwidth]{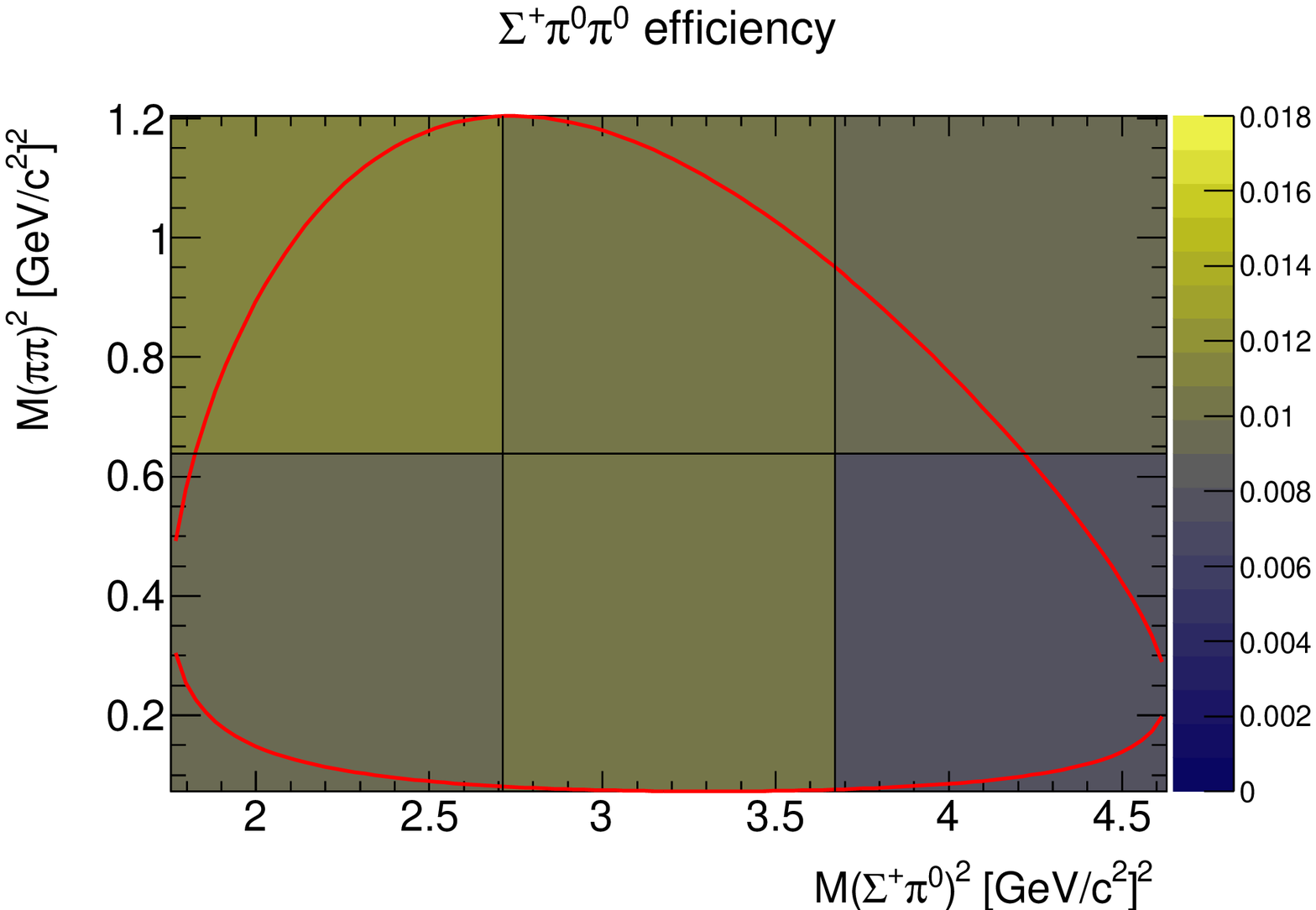}
     \put(20,19){\color{white}(a)}
   \put(45,19){(b)}
   \put(45,44){(c)}

  \end{overpic}

  \caption{Dalitz distribution binning and reconstruction efficiency in bins of $M(\Sigma^+\pi^0)^2$ vs.\ $M(\pi^0\pi^0)^2$ for the $\Lambda^+_c\to\Sigma^+\pi^0\pi^0$~channel. The curved line is the kinematic boundary of the Dalitz plot. The fit results in representative bins (a), (b) and (c) are shown in Fig.~\ref{fig:zero_fits}. \label{fig:zero}}
\end{figure}

\begin{figure}
  \includegraphics[width=0.48\textwidth]{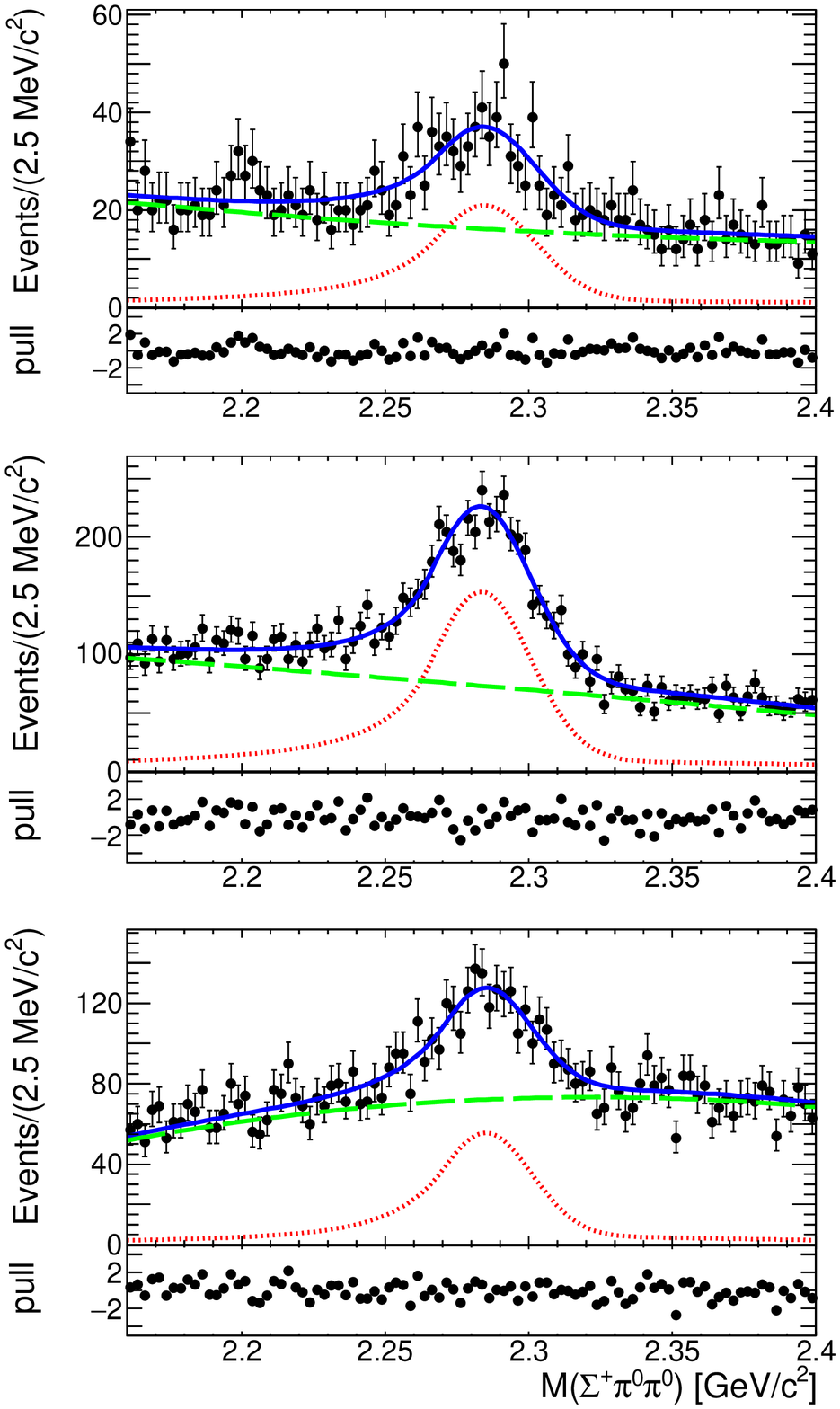}
  \caption{Fits (solid curves) in three representative Dalitz bins of the $\Lambda^+_c\to\Sigma^+\pi^0\pi^0$~channel. From top to bottom, the panels correspond to bins (a), (b) and (c) in Fig.~\ref{fig:zero}. The signal component is shown as the dotted curve, and the combinatorial background as the dashed curve. The pull distribution of the fit is shown at the bottom of each panel.} \label{fig:zero_fits}
\end{figure}

The signal yields in the $\Lambda^+_c\to pK^-\pi^+$, $\Sigma^0\pi^+\pi^0$, $\Sigma^+\pi^-\pi^+$, and $\Sigma^+\pi^0\pi^0$ modes are extracted using an unbinned extended maximum likelihood fit (EML)~\cite{extended} to the $\Lambda_c$-candidate invariant-mass distribution. The probability density functions (PDFs) of the signal and background models are typically defined between 2.2 and 2.4~GeV/$c^2$; for the $\Sigma^+\pi^0\pi^0$~mode, the lower bound is set to 2.14~GeV/$c^2$ to accommodate the longer signal tail at low invariant-masses. The signal in each channel is modeled by a combination of Gaussian, Breit-Wigner, and Crystal Ball~\cite{crystal ball} functions, sharing the same mean. Details are given in Table~\ref{tab:PDF}. The model is chosen empirically on MC samples and the width and peak position are in good agreement with data for all $\Sigma \pi \pi$ decay channels. For $pK^-\pi^+$, we find the signal shape to be 12\,\% broader in data. In the $\Sigma^0\pi^+\pi^0$ decay mode, $\Lambda_c \to \Lambda\pi^+\pi^0$ combined with one random photon causes a peak in the invariant-mass distribution that overlaps partially with the signal region. This background is included in the fit model. In all modes with a $\pi^0$ in the final state, $\pi^0$~candidates containing an incorrect photon produce a broad peak centered at the nominal $\Lambda_c^+$ mass. These self-cross-feed events, which amount to between 5\% and 23\% of true signal depending on the mode, are included in the signal component's PDF. For the combinatorial background, polynomials are used: cubic for $pK\pi$ {and $\Sigma^0\pi^+\pi^0$,} quadratic for other $\Sigma\pi\pi$ combinations.
\begin{table}
  \caption{Summary of the probability density functions (PDFs) used to model the signal component in the different $\Lambda_c^+$~modes. The alternative PDFs are used to estimate model uncertainties. A Gaussian function is abbreviated as `G', a Breit-Wigner function as `BW', and a Crystal Ball function as `CB'. The operator `$+$' denotes a linear sum of PDFs and `$\otimes$' stands for a convolution. All PDFs in the same decay channel share the same mean. The proportions of each function are determined from MC and fixed.}  \label{tab:PDF}
  \centering
  \begin{tabular}{ccc}
    \hline \hline
    $\Lambda_c^+$ mode  & PDF & Alternative PDF\\
    \hline
    $\Sigma^+\pi^+\pi^-$ & G $\otimes$ BW + G & G + G + BW  \\ 
    $\Sigma^0\pi^+\pi^0$ & CB + BW& CB + G  \\ 	
    $pK^-\pi^+$ & G + G + BW & G + G + G  \\
    $\Sigma^+\pi^0\pi^0$ & CB + G& CB + BW \\ 
    $\Lambda^0\pi^+\pi^0+\gamma$ & Bifurcated G + G & CB + G\\
    \hline\hline
  \end{tabular}
\end{table}
The reconstruction efficiency depends on the presence of intermediate resonances. To extract the signal yields in a model-independent way, the Dalitz distribution of each decay is binned and independent fits are performed in each bin. The binning and the Dalitz-bin efficiencies for $\Lambda_c^+\to pK^-\pi^+$, $\Sigma^0\pi^+\pi^0$, $\Sigma^+\pi^+\pi^-$, and $\Sigma^+\pi^0\pi^0$ are shown in Figs.~\ref{fig:pkpi}, \ref{fig:Sig0}, \ref{fig:Sigp} and \ref{fig:zero}, respectively. The PDF parameters in each bin are determined from simulation. In the fit to $\Sigma \pi \pi$ real data, only the normalizations of the signal and combinatorial background are floated, except in the $\Lambda^+_c\to\Sigma^0\pi^+\pi^0$~channel, where the distinct contribution of the $\Lambda\pi^+\pi^0+\gamma$ background is { also} determined bin by bin. For $\Lambda_c^+\to pK^-\pi^+$, both the background polynomial and the width of the signal component are allowed to float. {For $\Sigma \pi \pi$, the width is measured on the full sample and fixed for yield extraction.} Figures~\ref{fig:pkpi_fits}, \ref{fig:Sig0_fits}, \ref{fig:Sigp_fits} and \ref{fig:zero_fits} show sample Dalitz-bin plots to illustrate the extraction of the signal yields.

At the next step, the extracted yields in each bin are efficiency-corrected and summed over the Dalitz plot  to give the total yield
\begin{equation}
  y = \sum\limits_{i} \frac{y_i}{\epsilon_i}~. \label{equ:eff}
\end{equation}
Here, the index $i$ runs over the Dalitz plot bins shown in Figs.~\ref{fig:pkpi}, \ref{fig:Sig0}, \ref{fig:Sigp} and \ref{fig:zero}, and $y_i$ and $\epsilon_i$ are the extracted signal yield and the reconstruction efficiency, respectively, for bin~$i$. The result for the total efficiency-corrected signal yield $y$ is given for each mode in Table~\ref{tab:paperstamp}.
\begin{table}
  \caption{Efficiency-corrected signal yields for the different $\Lambda_c^+$~modes in multiples of $10^3$. The quoted error is the quadratic sum of the yield uncertainty from the fit in the individual Dalitz bins.} \label{tab:paperstamp}
  \centering
 \begin{tabular}{cc}
   \hline \hline
   Final state  & $\sum_i{y_i/\epsilon_i}$ $[\times10^3]$\\
   \hline
   $\Sigma^+\pi^-\pi^+$ & $ 2687  \pm 10  $  \\
   $\Sigma^0\pi^+\pi^0$ & $2661 \pm 24 $ \\
   $pK^-\pi^+$ & $7249  \pm 9 $ \\
   $\Sigma^+\pi^0\pi^0$ & $ 925 \pm 22 $ \\
   \hline\hline
 \end{tabular}
\end{table}
\section{Results and systematic uncertainty}

The branching fractions of the decays $\Lambda^+_c\to\Sigma^+\pi^-\pi^+$, $\Lambda^+_c\to\Sigma^0\pi^+\pi^0$, and $\Lambda^+_c\to\Sigma^+\pi^0\pi^0$ relative to that of the decay $\Lambda^+_c\to pK^-\pi^+$ are calculated from the total efficiency-corrected signal yields given in Table~\ref{tab:paperstamp}:
\begin{equation}
  \frac{{\cal B}(\Lambda_c\to \Sigma \pi \pi)}{{\cal B}(\Lambda_c^+\to pK^-\pi^+)} = \frac{y_{\Sigma \pi \pi}}{y_{p K \pi}{\cal B}_\mathrm{PDG}}~. \label{equ:br}
\end{equation}
Here, ${\cal B}_\mathrm{PDG}$ denotes the subdecay branching fractions of $\Sigma^+$ and $\Lambda$~\cite{PDG}. All results are summarized in Table~\ref{tab:brresult}.
\begin{table*}
  \caption{Branching-fraction values determined by this analysis. The second column gives the branching fractions of the decays $\Lambda^+_c\to\Sigma^+\pi^-\pi^+$, $\Lambda^+_c\to\Sigma^0\pi^+\pi^0$, and $\Lambda^+_c\to\Sigma^+\pi^0\pi^0$ relative to the branching fraction of the decay $\Lambda^+_c\to pK^-\pi^+$. The third column lists the absolute branching fractions taking ${\cal B}(\Lambda^+_c\to pK^-\pi^+)=6.35\pm 0.33$~\cite{PDG}. Errors are statistical, systematic, and from ${\cal B}(pK\pi)$, respectively. In the final column, the current world average is given.} \label{tab:brresult}
  \centering
  \addtolength{\tabcolsep}{2pt}
  \begin{tabular}{cccc}
    \hline \hline
    Final state & ${\cal B}(\Sigma\pi\pi)/{\cal B}(pK\pi)$ & ${\cal B}(\Sigma\pi\pi)$ [\%] & ${\cal B}_{WA}(\Sigma\pi\pi)$ [\%] \\
\hline
    $\Sigma^+\pi^-\pi^+$ &  $0.7{19}\pm 0.003\pm 0.024$ & $4.{57}\pm 0.02\pm 0.1{5}\pm 0.2{4}$ & $4.57\pm0.29$\\
    $\Sigma^0\pi^+\pi^0$ & $0.{575}\pm 0.005\pm 0.0{36}$ & $3.{65}\pm 0.03\pm 0.{23}\pm 0.1{9}$ &$2.3\pm0.9 $\\
    $\Sigma^+\pi^0\pi^0$ & $0.{247}\pm 0.006\pm 0.01{9}$ & $1.{57}\pm 0.04\pm 0.1{2}\pm 0.0{8}$ & -\\
    \hline\hline		
  \end{tabular}  
  \addtolength{\tabcolsep}{-2pt}
\end{table*}

The following uncertainties are taken into account and listed in Table~\ref{tab:sys}.
Unless stated otherwise, we assume no correlation in the individual systematic error components and so add them in quadrature.
The systematic uncertainty related to the pion and kaon identification efficiency is estimated from kinematically identified $D^{*+}\to D^0 \pi^+$, $D^0\to K^- \pi^+$ real-data events. These events are used both to derive a correction to the MC simulation and to determine the systematic uncertainties of pion and kaon identification. All channels except $\Sigma^+\pi^0\pi^0$ include a charged pion, directly produced in the $\Lambda_c^+$ decay. The uncertainty caused by the PID selection of this particle cancels in the ratio.
The uncertainty introduced by proton identification is determined from the ratio of yields of the decay $\Lambda\to p\pi$ with and without the proton identification requirement. The difference in the ratio between MC and data is used to correct the efficiency; the statistical uncertainty is treated as a systematic error.
The systematic uncertainty due to $\Lambda$ reconstruction is estimated by considering the data--MC difference of tracks displaced from the IP, the $\Lambda$ proper time, and $\Lambda$ mass distributions. The weighted average over the momentum range is taken as the total uncertainty.
A study of $\tau^-\to\pi^-\pi^0\nu_\tau$ decays described in~\cite{pi0} is used to correct for MC--data discrepancies in the $\pi^0$ reconstruction efficiency.
We check model uncertainties by varying the PDF parameters fixed from MC within their statistical uncertainties and repeat the fits one thousand times for each bin. The change in the central value plus the width of the distribution, in terms of standard deviation, of fit results is taken as a systematic error in a given bin and the weighted sum is taken as the total systematic error. Furthermore, we use alternate signal PDFs as described in Table~\ref{tab:PDF} and alternate background PDFs whose polynomial order is increased by one.
The residual Dalitz model dependence of our fitting method is checked by repeating the fit with a four times finer binning. The difference in the yields is taken as a systematic error. Limited statistics preclude us from using a finer binning in the case of $\Sigma^+\pi^0\pi^0$. Here, we compare the efficiency-corrected signal yield with the fit on the unbinned sample and take the difference as a systematic error. 
The uncertainty due to tracking is 0.35\,\% per charged track. We only apply this uncertainty to $pK^-\pi^+$ in the ratio with $\Sigma^+\pi^0\pi^0$. In the other decay modes, the equal number of charged tracks in the measured and reference modes causes this uncertainty to cancel.
For the reconstruction of the photon from the $\Sigma^0\to \Lambda \gamma$ decay, we apply half the uncertainty for low-momentum (below 200~MeV/$c$) $\pi^0$ reconstruction. The additional uncertainty compared to general $\pi^0$ reconstruction is obtained from a study of $B^0 \to D^{\ast-}\pi^+$ and $B^+ \to D^{\ast0}\pi^+$ decays to determine the data--MC ratio in bins of pion momentum from the $D^{\ast}$ decay. The overall systematic error is obtained by linear summation of this uncertainty and the results of the $\tau^-\to\pi^-\pi^0\nu_\tau$ study mentioned previously. {Possible uncertainties introduced by the BDT selector are studied by loosening the selection as far as possible while maintaining a plausible fit quality. The changes in the efficiency-corrected  yields are found to be consistent with zero within the statistical uncertainty.}

\begin{table*}
  \caption{Summary of the relative systematic error contributions to efficiency-corrected signal yields (in \%). Only uncertainties that do not cancel in the branching-fraction ratios are given. For $pK^-\pi^+$, the cancellation of uncertainties with $\Sigma^+\pi^-\pi^+$ and $\Sigma^0\pi^+\pi^0$ or ($\Sigma^+\pi^0\pi^0$) is taken into account.} \label{tab:sys}
  \centering
  \begin{tabular}{ccccc}
    \hline \hline
    Source &
    $\Sigma^+\pi^+\pi^-$ & $\Sigma^+\pi^0\pi^0$ & $\Sigma^0\pi^+\pi^0$ & $pK^-\pi^+$\\
    \hline
    K $\pi$ identification  & 1.16& - &1.88& 1.18(1.64)\\
    Proton identification& 0.42 & 0.39 & 0.39 &0.47\\
    $\Lambda$ identification & $-$ & $-$ & 2.68& $-$\\
    $\pi^0$ identification & 2.44 & 6.82 & 2.27 & $-$\\
    PDF model  &0.6 & 2.18 &  3.13&1.04\\ 
    Dalitz structure  & 0.0  &0.06 &0.71  &0\\ 
    Tracking & 0& 0& 0& 0(0.7)\\
    $\gamma$ identification & 0& 0& 3.15& 0\\
    MC statistics & 0.1 & 0.6 & 0.3 &0\\ 
    $\cal{B}_\mathrm{PDG}$ & 0.3 & 0.3 & 0.5 & $-$\\
    \hline
    Total & 2.82 & 7.20  & 5.98 & 1.65
    (2.13)\\
    \hline \hline
  \end{tabular}
\end{table*}

\section{Summary}

We analyze the decays $\Lambda^+_c\to\Sigma^+\pi^-\pi^+$, $\Lambda^+_c\to\Sigma^0\pi^+\pi^0$, and $\Lambda^+_c\to\Sigma^+\pi^0\pi^0$ using the full Belle data set at or near the $\Upsilon(4S)$~resonance. Using a model-independent approach, we fit the signal yields in separate bins of the decay Dalitz distribution to avoid uncertainties introduced by intermediate resonances. We measure branching-fraction ratios of
 
$$\frac{{\cal B}(\Lambda_c^+\rightarrow\Sigma^+\pi^-\pi^+)}{{\cal B}(\Lambda^+_c\to pK^-\pi^+)} = 0.7{19}\pm 0.003\pm 0.02{4},$$
$$\frac{{\cal B}(\Lambda_c^+\rightarrow\Sigma^0\pi^+\pi^0)}{{\cal B}(\Lambda^+_c\to pK^-\pi^+)} = 0.{575}\pm 0.005\pm 0.0{36},  $$
$$\frac{{\cal B}(\Lambda_c^+\rightarrow\Sigma^+\pi^0\pi^0)}{{\cal B}(\Lambda^+_c\to pK^-\pi^+)} = 0.{247}\pm 0.006\pm 0.01{9}. $$
 \newline

The first (second) quoted uncertainties are statistical (systematic).
Assuming ${\cal B}(\Lambda^+_c\to pK^-\pi^+)=6.35\pm 0.33$~\cite{PDG}, we obtain
 $${\cal B}(\Lambda_c^+\!\rightarrow\Sigma^+\pi^-\pi^+) = 4.{57}\pm 0.02\pm 0.1{5}\pm 0.2{4}\,\%,$$
$${\cal B}(\Lambda_c^+\!\rightarrow\Sigma^0\pi^+\pi^0) = 3.{65}\pm 0.03\pm 0.{23}\pm 0.1{9}\,\%, $$
$${\cal B}(\Lambda_c^+\!\rightarrow\Sigma^+\pi^0\pi^0) = 1.{57}\pm 0.04\pm 0.1{2}\pm 0.0{8}\,\%.$$
The third quoted uncertainties are due to ${\cal B}(pK^-\pi^+)$.
The results agree with previous experimental findings~\cite{pkpi_BESIII,CLEO_Sig0} where they exist. This is the first measurement of $\Lambda_c^+\rightarrow\Sigma^+\pi^0\pi^0$. The measurement of $\Lambda_c^+\rightarrow\Sigma^0\pi^+\pi^0$ is four times more precise than the current world average.

\begin{acknowledgments}
We thank the KEKB group for the excellent operation of the
accelerator; the KEK cryogenics group for the efficient
operation of the solenoid; and the KEK computer group,
the National Institute of Informatics, and the 
PNNL/EMSL computing group for valuable computing
and SINET5 network support.  We acknowledge support from
the Ministry of Education, Culture, Sports, Science, and
Technology (MEXT) of Japan, the Japan Society for the 
Promotion of Science (JSPS), and the Tau-Lepton Physics 
Research Center of Nagoya University; 
the Australian Research Council;
Austrian Science Fund under Grant No.~P 26794-N20 and Doctoral program No. W1252-N27;
the National Natural Science Foundation of China under Contracts 
No.~10575109, No.~10775142, No.~10875115, No.~11175187, No.~11475187, 
No.~11521505 and No.~11575017;
the Chinese Academy of Science Center for Excellence in Particle Physics; 
the Ministry of Education, Youth and Sports of the Czech
Republic under Contract No.~LTT17020;
the Carl Zeiss Foundation, the Deutsche Forschungsgemeinschaft, the
Excellence Cluster Universe, and the VolkswagenStiftung;
the Department of Science and Technology of India; 
the Istituto Nazionale di Fisica Nucleare of Italy; 
National Research Foundation (NRF) of Korea Grants No.~2014R1A2A2A01005286, No.~2015R1A2A2A01003280,
No.~2015H1A2A1033649, No.~2016R1D1A1B01010135, No.~2016K1A3A7A09005603, No.~2016R1D1A1B02012900; Radiation Science Research Institute, Foreign Large-size Research Facility Application Supporting project and the Global Science Experimental Data Hub Center of the Korea Institute of Science and Technology Information;
the Polish Ministry of Science and Higher Education and 
the National Science Center;
the Ministry of Education and Science of the Russian Federation and
the Russian Foundation for Basic Research;
the Slovenian Research Agency;
Ikerbasque, Basque Foundation for Science and
MINECO (Juan de la Cierva), Spain;
the Swiss National Science Foundation; 
the Ministry of Education and the Ministry of Science and Technology of Taiwan;
and the U.S.\ Department of Energy and the National Science Foundation.
\end{acknowledgments}


\begin{thebibliography}{99}
\bibitem{cc}
  Throughout this paper, the inclusion of charge-conjugate states is implied unless stated otherwise. For $\Sigma$-hyperons this refers to their respective antiparticles.
\end{thebibliography}
\end{document}